\documentclass[twocolumn,showpacs,preprintnumbers,aps,amsmath,amssymb]{revtex4}
\begin{document}
%
\bibliographystyle{try}

\topmargin 0.1cm

\newcounter{univ_counter}
\setcounter{univ_counter} {0}

\addtocounter{univ_counter} {1} 
\edef\OHIOU{$^{\arabic{univ_counter}}$ } 

\addtocounter{univ_counter} {1} 
\edef\JLAB{$^{\arabic{univ_counter}}$ } 

\addtocounter{univ_counter} {1} 
\edef\VIRGINIA{$^{\arabic{univ_counter}}$ } 

\addtocounter{univ_counter} {1} 
\edef\FIU{$^{\arabic{univ_counter}}$ } 

\addtocounter{univ_counter} {1} 
\edef\ASU{$^{\arabic{univ_counter}}$ } 

\addtocounter{univ_counter} {1} 
\edef\SACLAY{$^{\arabic{univ_counter}}$ } 

\addtocounter{univ_counter} {1} 
\edef\UCLA{$^{\arabic{univ_counter}}$ } 

\addtocounter{univ_counter} {1} 
\edef\CMU{$^{\arabic{univ_counter}}$ } 

\addtocounter{univ_counter} {1} 
\edef\CUA{$^{\arabic{univ_counter}}$ } 

\addtocounter{univ_counter} {1} 
\edef\CNU{$^{\arabic{univ_counter}}$ } 

\addtocounter{univ_counter} {1} 
\edef\UCONN{$^{\arabic{univ_counter}}$ } 

\addtocounter{univ_counter} {1} 
\edef\DUKE{$^{\arabic{univ_counter}}$ } 

\addtocounter{univ_counter} {1} 
\edef\EDINBURGH{$^{\arabic{univ_counter}}$ } 

\addtocounter{univ_counter} {1} 
\edef\FSU{$^{\arabic{univ_counter}}$ } 

\addtocounter{univ_counter} {1} 
\edef\GWU{$^{\arabic{univ_counter}}$ } 

\addtocounter{univ_counter} {1} 
\edef\GLASGOW{$^{\arabic{univ_counter}}$ } 

\addtocounter{univ_counter} {1} 
\edef\INDO{$^{\arabic{univ_counter}}$ } 

\addtocounter{univ_counter} {1} 
\edef\INFNFR{$^{\arabic{univ_counter}}$ } 

\addtocounter{univ_counter} {1} 
\edef\INFNGE{$^{\arabic{univ_counter}}$ } 

\addtocounter{univ_counter} {1} 
\edef\ORSAY{$^{\arabic{univ_counter}}$ } 

\addtocounter{univ_counter} {1} 
\edef\ITEP{$^{\arabic{univ_counter}}$ } 

\addtocounter{univ_counter} {1} 
\edef\JMU{$^{\arabic{univ_counter}}$ } 

\addtocounter{univ_counter} {1} 
\edef\KYUNGPOOK{$^{\arabic{univ_counter}}$ } 

\addtocounter{univ_counter} {1} 
\edef\MIT{$^{\arabic{univ_counter}}$ } 

\addtocounter{univ_counter} {1} 
\edef\UMASS{$^{\arabic{univ_counter}}$ } 

\addtocounter{univ_counter} {1} 
\edef\MSU{$^{\arabic{univ_counter}}$ } 

\addtocounter{univ_counter} {1} 
\edef\UNH{$^{\arabic{univ_counter}}$ } 

\addtocounter{univ_counter} {1} 
\edef\NSU{$^{\arabic{univ_counter}}$ } 

\addtocounter{univ_counter} {1} 
\edef\ODU{$^{\arabic{univ_counter}}$ } 

\addtocounter{univ_counter} {1} 
\edef\PITT{$^{\arabic{univ_counter}}$ } 

\addtocounter{univ_counter} {1} 
\edef\RPI{$^{\arabic{univ_counter}}$ } 

\addtocounter{univ_counter} {1} 
\edef\RICE{$^{\arabic{univ_counter}}$ } 

\addtocounter{univ_counter} {1} 
\edef\URICH{$^{\arabic{univ_counter}}$ } 

\addtocounter{univ_counter} {1} 
\edef\SCAROLINA{$^{\arabic{univ_counter}}$ } 

\addtocounter{univ_counter} {1} 
\edef\UTEP{$^{\arabic{univ_counter}}$ } 

\addtocounter{univ_counter} {1} 
\edef\VT{$^{\arabic{univ_counter}}$ } 

\addtocounter{univ_counter} {1} 
\edef\WM{$^{\arabic{univ_counter}}$ } 

\addtocounter{univ_counter} {1} 
\edef\YEREVAN{$^{\arabic{univ_counter}}$ } 

\title{First Measurement of Transferred Polarization in the Exclusive
$\vec{e} p \to e' K^+ \vec{\Lambda}$ Reaction}


 \author{ 
D.S.~Carman,\OHIOU\
K.~Joo,\JLAB$\!\!^,$\VIRGINIA\
M.D.~Mestayer,\JLAB\
B.A.~Raue,\FIU$\!\!^,$\JLAB\
G.~Adams,\RPI\
P.~Ambrozewicz,\FIU\
E.~Anciant,\SACLAY\
M.~Anghinolfi,\INFNGE\
D.S.~Armstrong,\WM\
B.~Asavapibhop,\UMASS
% ...
\ 
\\
(CLAS collaboration)
} 

\affiliation{\OHIOU Ohio University, Athens, Ohio  45701}
\affiliation{\JLAB Thomas Jefferson National Accelerator Laboratory, Newport News, Virginia 23606}
\affiliation{\VIRGINIA University of Virginia, Charlottesville, Virginia 22901}
\affiliation{\FIU Florida International University, Miami, Florida 33199}
\affiliation{\ASU Arizona State University, Tempe, Arizona 85287}
\affiliation{\SACLAY CEA-Saclay, DAPNIA-SPhN, F91191 Gif-sur-Yvette Cedex, France}
\affiliation{\UCLA University of California at Los Angeles, Los Angeles, California  90095}
\affiliation{\CMU Carnegie Mellon University, Pittsburgh, Pennsylvania 15213}
\affiliation{\CUA Catholic University of America, Washington, D.C. 20064}
\affiliation{\CNU Christopher Newport University, Newport News, Virginia 23606}
\affiliation{\UCONN University of Connecticut, Storrs, Connecticut 06269}
\affiliation{\DUKE Duke University, Durham, North Carolina 27708}
\affiliation{\EDINBURGH Edinburgh University, Edinburgh EH9 3JZ, United Kingdom}
\affiliation{\FSU Florida State University, Tallahassee, Florida 32306}
\affiliation{\GWU The George Washington University, Washington, DC 20052}
\affiliation{\GLASGOW University of Glasgow, Glasgow G12 8QQ, United Kingdom}
\affiliation{\INDO Jurusan Fisika, FMIPA, Universitas Indonesia, Depok 16424, Indonesia}
\affiliation{\INFNFR INFN, Laboratori Nazionali di Frascati, P.O. 13,00044 Frascati, Italy}
\affiliation{\INFNGE INFN, Sezione di Genova and Dipartimento di Fisica, Universit\`a di Genova, 16146 Genova, Italy}
\affiliation{\ORSAY Institut de Physique Nucleaire d'ORSAY, IN2P3, BP1, 91406 Orsay, France}
\affiliation{\ITEP Institute of Theoretical and Experimental Physics, Moscow, 117259, Russia}
\affiliation{\JMU James Madison University, Harrisonburg, Virginia 22807}
\affiliation{\KYUNGPOOK Kyungpook National University, Daegu 702-701, South Korea}
\affiliation{\MIT Massachusetts Institute of Technology, Cambridge, Massachusetts  02139}
\affiliation{\UMASS University of Massachusetts, Amherst, Massachusetts  01003}
\affiliation{\MSU Moscow State University, 119899 Moscow, Russia}
\affiliation{\UNH University of New Hampshire, Durham, New Hampshire 03824}
\affiliation{\NSU Norfolk State University, Norfolk, Virginia 23504}
\affiliation{\ODU Old Dominion University, Norfolk, Virginia 23529}
\affiliation{\PITT University of Pittsburgh, Pittsburgh, Pennsylvania 15260}
\affiliation{\RPI Rensselaer Polytechnic Institute, Troy, New York 12180}
\affiliation{\RICE Rice University, Houston, Texas 77005}
\affiliation{\URICH University of Richmond, Richmond, Virginia 23173}
\affiliation{\SCAROLINA University of South Carolina, Columbia, South Carolina 29208}
\affiliation{\UTEP University of Texas at El Paso, El Paso, Texas 79968}
\affiliation{\VT Virginia Polytechnic Institute and State University, Blacksburg, Virginia 24061}
\affiliation{\WM College of William and Mary, Williamsburg, Virginia 23187}
\affiliation{\YEREVAN Yerevan Physics Institute, 375036 Yerevan, Armenia}

\date{\today}

\begin{abstract}

The first measurements of the transferred polarization for the exclusive
$\vec{e}p \to e'K^+ \vec{\Lambda}$ reaction have been performed in Hall B 
at the Thomas Jefferson National Accelerator Facility using the CLAS 
spectrometer.  A 2.567~GeV electron beam was used to measure the hyperon 
polarization over a range of $Q^2$ from 0.3 to 1.5~(GeV/c)$^2$, $W$ from 
1.6 to 2.15~GeV, and over the full center-of-mass angular range of the $K^+$ 
meson.  Comparison with predictions of hadrodynamic models indicates strong 
sensitivity to the underlying resonance contributions.  A non-relativistic
quark model interpretation of our data suggests that the $s \bar{s}$ quark
pair is produced with spins predominantly anti-aligned.  Implications for
the validity of the widely used $^3\!P_0$ quark-pair creation operator are
discussed.
\end{abstract}


\maketitle
\newpage

We present here the first measurements of spin transfer in the 
nucleon resonance region from a longitudinally polarized electron beam to
the $\Lambda$ hyperon produced in the exclusive $p(\vec{e},e'K^+)\vec{\Lambda}$
reaction.  Understanding nucleon resonance excitation continues to provide a 
major challenge to hadronic physics due to the non-perturbative nature of QCD 
at these energies.  Studies of strange final states have the potential to 
uncover baryonic resonances that do not couple or couple only weakly to the 
$\pi N$ channel due to the different hadronic vertices.  Recent symmetric 
quark-model calculations predict more states than have been found experimentally
\cite{capstick}.  The issue of whether these missing resonances do in fact exist 
is directly tied to the question of whether certain quark degrees of freedom might 
be ``frozen out'' as in, e.g., certain diquark models~\cite{alkofer}.  This 
question is central to our understanding of baryon structure.    

In the absence of direct QCD predictions, the theoretical framework 
involving hadrodynamic models has been extensively applied to the study 
of electromagnetic production of pseudoscalar mesons
\cite{wjc,mart,saghai,janssen}.  Considerable effort over the last two 
decades has been expended to develop these effective Lagrangian models; however,
their predictive powers are still limited by a sparsity of data.  Model 
fits to the existing cross section data are generally obtained at the expense of 
many free parameters, and these unpolarized data alone are not 
sufficiently sensitive to fully understand the reaction mechanism as they probe 
only a small portion of the full response.  Our double-polarization data can 
provide significant new constraints on the basic parameters of these models, 
increasing their discriminatory power and allowing for a quantitative measure of 
whether or not new ``missing'' resonances might be required to explain these and 
other hyperon production data.  

Alternatively, our data provide interesting, and perhaps surprising, 
information about the nature of quark-pair production.  There is a growing 
body of evidence that the appropriate degrees of freedom to describe the 
phenomenology of hadronic decays are constituent quarks held together by a 
gluonic flux-tube~\cite{isgur}.  
The non-perturbative nature of the flux-tube gives rise to the well-known linear
potential of heavy-quark confinement ($dV/dr \sim 1$~GeV/fm)~\cite{bali}.  Other 
properties of the flux-tube can be determined by studying $q \bar{q}$ pair 
production, since this is widely believed to produce the color field 
neutralization that actually breaks the flux-tube.  Since the suggestions in the 
1970's that a quark pair with vacuum quantum numbers is responsible for breaking 
the color flux-tube (the $^3\!P_0$ model \cite{leyaouanc}), theorists 
have used increasingly sophisticated and realistic models to understand the 
experimental data~\cite{ackleh,geiger}.

The most sensitive experimental tests to date have measured the ratio of strong 
amplitudes differing in their orbital angular momenta in certain meson decays  
\cite{geiger}. Since the $^3\!P_0$ operator has $S$=1 and $L$=1, it implies a 
different amplitude ratio than, e.g., a $^3\!S_1$ operator with $S$=1 and $L$=0,
corresponding to one gluon exchange.  Later, we will argue that the 
spin properties of the quark-pair creation operator might be responsible for the 
observed trends in the $\Lambda$ polarization.  Furthermore, by analogy of
an argument used to explain the observed transverse polarization of 
$\Lambda$ hyperons in exclusive $pp \to p K^+ \Lambda$~\cite{boros}, we conclude 
that the relevant quark-pair creation operator dominating our reaction
produces the $s\bar{s}$ pair with spins anti-aligned.  This finding, if 
confirmed by further calculations, brings into 
question the universal applicability of the $^3\!P_0$ model.  This has important 
implications since many, if not most, calculations of hadronic spectroscopy use 
the $^3\!P_0$ operator to calculate the transition to the final-state particles
\cite{barnes}.

The Continuous Electron Beam Accelerator Facility (CEBAF) at Jefferson 
Laboratory provides multi-GeV electron beams with longitudinal polarization 
up to 80\%.  In Hall B of this facility is the CEBAF Large Acceptance 
Spectrometer (CLAS)~\cite{smith}, a detector constructed around six 
superconducting coils that generate a toroidal magnetic field to momentum-analyze 
charged particles.  The detection 
system consists of multiple layers of drift chambers to determine charged-particle 
trajectories, {\v C}erenkov detectors for electron/pion 
separation, scintillation counters for flight-time measurements, and 
calorimeters to identify electrons and high-energy neutral particles. Operating 
luminosity with the unpolarized liquid-H$_2$ target is 
$\sim$1$\times$10$^{34}$~cm$^{-2}$ sec$^{-1}$.

\begin{figure}[htbp]
\vspace{2.7cm}
\includegraphics{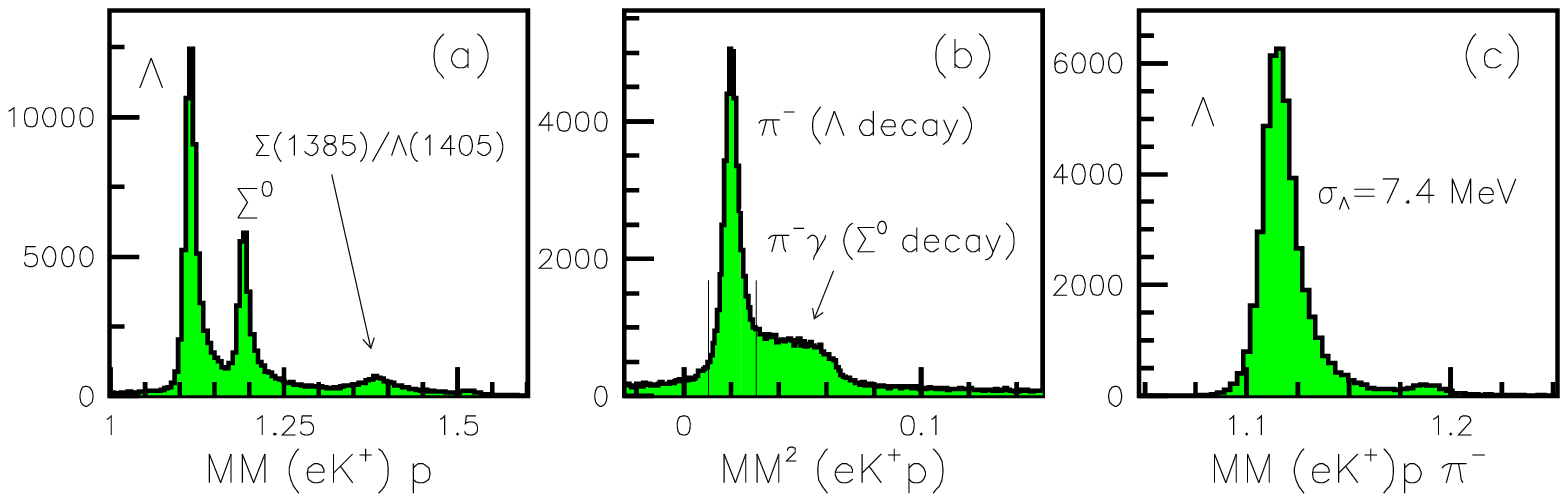}
\caption[]{Missing-mass spectra (GeV) for the reactions (a) $p(e,e'K^+)X$ 
and (b) $p(e,e'K^+p)X$. (c) The hyperon distribution after cutting on the 
low-mass peak in (b).  Data from 2.567~GeV CLAS data summing over all 
$Q^2$ and $W$.} 
\label{hypfs}
\end{figure}

The large acceptance of CLAS has enabled us to detect the final-state electron and 
kaon, and the proton from the decay of the $\Lambda$ hyperon at a beam energy 
of 2.567~GeV, over a range of momentum transfer $Q^2$ from 0.3 to 1.5~(GeV/c)$^2$ 
and invariant energy $W$ from 1.6 to 2.15~GeV, while providing full angular 
coverage in the kaon center-of-mass (CM).
Hyperon identification with CLAS relies on missing-mass reconstructions.  
Fig.~\ref{hypfs}a shows the missing-mass for $p(e,e'K^+)X$ where a
proton has also been detected.  Fig.~\ref{hypfs}b 
shows the missing mass for $p(e,e'K^+p)X$, where the final-state 
proton can come from the decay of the $\Lambda$(1115) (missing $\pi^-$) or 
the $\Sigma^0$(1192) (missing $\pi^- \gamma$).  Fig.~\ref{hypfs}c shows the 
resulting hyperon spectrum after a cut on the $\pi^-$ peak in Fig.~\ref{hypfs}b.

\begin{figure}[htbp]
\vspace{2.8cm}
\includegraphics{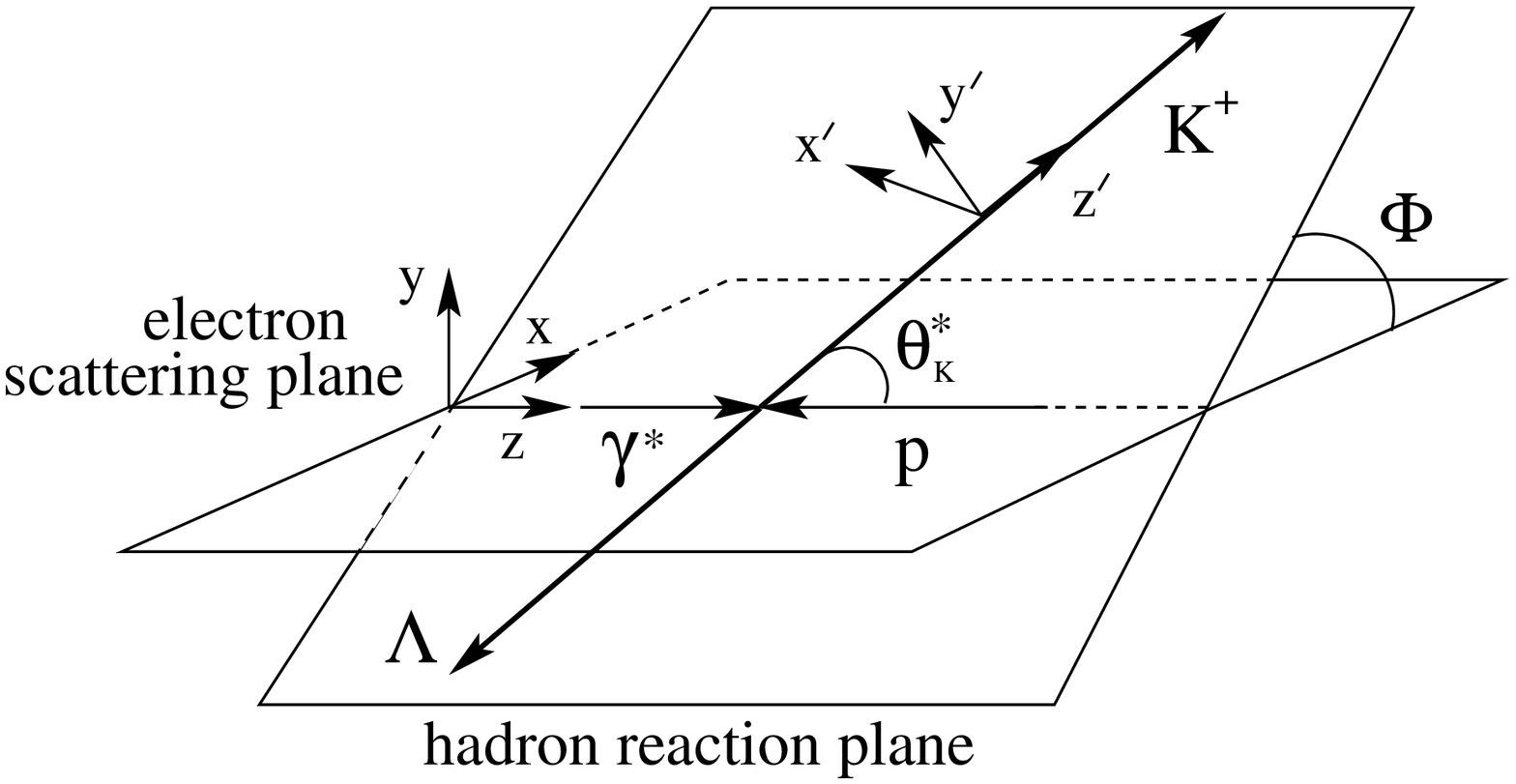}
\caption[]{Center-of-mass coordinate system highlighting the definitions
of the different spin-quantization axis choices for the 
final-state $\Lambda$ hyperon used in the polarization analysis.}
\label{coor7}
\end{figure}

An attractive feature of the mesonic decay $\Lambda \to p\pi^-$ comes from 
its self-analyzing nature.  This weak decay has an asymmetric angular 
distribution with respect to the $\Lambda$ spin direction due to an 
interference between parity non-conserving ($s$-wave) and parity-conserving 
($p$-wave) amplitudes.  The decay-proton distribution in the $\Lambda$ rest 
frame (RF) for each beam helicity state is of the form:
\begin{equation}
\label{polform}
\frac{dN^{\pm}}{d\cos \theta_p^{RF}} = N [1 + {\alpha} (P^0 \pm P_b P') \cos \theta_p^{RF}],
\end{equation}
\noindent
where $P_b$ is the average beam polarization and $\alpha$=0.642$\pm$0.013 is the 
weak decay asymmetry parameter~\cite{pdg}.  The $\Lambda$ polarization is the 
sum of $P^0$, the induced polarization, and $P'$, the helicity-dependent 
transferred polarization, both defined with respect to a particular set of 
spin-quantization axes.  This latter quantity is the focus of this work. 
Fig.~\ref{coor7} highlights two standard choices for the spin-quantization axes 
defined in relation to the electron and hadron reaction planes.  

Using eq.(\ref{polform}), we can express the acceptance-corrected
yield asymmetries in terms of the average transferred polarization for
each kinematic bin as:

\begin{equation}
\label{asymmetry}
A_{\xi}(\cos \theta_p^{RF}) = \frac{N^+_{\xi} - N^-_{\xi}}{N^+_{\xi} + N^-_{\xi}} 
= \frac{\alpha P_b \cos \theta_p^{RF} P_{\xi}'}{1 + \alpha P^0_{\xi} 
\cos \theta_p^{RF}}.
\end{equation}
\noindent
Here $N^{\pm}_{\xi}(\cos \theta_p^{RF})$ are the decay-proton helicity-gated 
yields with respect to the different spin-quantization axes 
$\xi = \hat{i},\hat{j},\hat{k}$.  Using this method we are quite insensitive 
to the form of the CLAS acceptance function.  

Using the notation of Ref.~\cite{knochlein}, the most general form for the
virtual photo-absorption CM cross section for the $p(\vec{e},e'K^+)\vec{\Lambda}$ 
reaction from an unpolarized proton target, allowing for both a 
polarized-electron beam and recoil hyperon is given by:
\begin{eqnarray}
\label{csec1}
\frac{d \sigma_v}{d\Omega_K^*}\!\!\! &=& \!\!\! {\cal{K}} 
   \!\!\!\!\!\!\!\sum_{\beta=0,x',y',z'} \!\!\!\!\!
   \Bigl( R_T^{\beta 0} + \epsilon_L R_L^{\beta 0} + c_+(^c\!R_{LT}^{\beta 0} 
   \cos \Phi \nonumber \\
&+& \!\! ^s\!\!R_{LT}^{\beta 0} \sin \Phi) + \epsilon (^c\!R_{TT}^{\beta 0} \cos 2 \Phi + 
^s\!\!R_{TT}^{\beta 0} \sin 2 \Phi) \nonumber \\
+ P_b \!\!\!&c_-& \!\!\!(^c\!R_{LT'}^{\beta 0} \cos \Phi + ^s\!\!R_{LT'}^{\beta 0}\sin \Phi)
+ P_b c_0 R_{TT'}^{\beta 0} \Bigr).
\end{eqnarray}
\noindent
The $R_i$ are the transverse, longitudinal, and interference response 
functions that relate to the underlying hadronic current and implicitly 
contain the $\Lambda$ polarization.  The sum over $\beta$ includes 
contributions from the polarization with respect to the $(x',y',z')$ axes 
(see Fig. \ref{coor7}), and the $\beta=0$ terms account for the unpolarized 
response.  Here $c_{\pm} = \sqrt{2 \epsilon_L (1 \pm \epsilon)}$ and 
$c_0 = \sqrt{1- \epsilon^2}$, where $\epsilon$
($\epsilon_L=\epsilon Q^2/(k_{\gamma}^{CM})^2$) is the 
transverse (longitudinal) polarization of the virtual photon, 
${\cal{K}}= \vert \vec{q}_K\vert / k_{\gamma}^{CM}$, and $\Phi$ is the 
angle between the electron and hadron planes.  The $c$ and $s$ labels
indicate whether $R_i$ multiplies a sine or cosine term.

From the cross section of eq.(\ref{csec1}), the induced and transferred 
polarization components in the $(x',y',z')$ system are given by~\cite{schmieden}:
\begin{eqnarray}
\label{pind1}
\sigma_0 P_{\xi}^0 \!\!\!&=&\!\!\! {\cal{K}} ( c_+ R_{LT}^{\xi 0}\sin{\Phi} 
      + \epsilon\ R_{TT}^{\xi 0}\sin{2\Phi} ),~~~~ \xi=x'\!,z'~~~\\
\sigma_0 P_{y'}^0 \!\!\!&=&\!\!\! {\cal{K}} ( R_T^{y'0} \!+\! \epsilon_L R_L^{y'0}
      \!+\! c_+ R_{LT}^{y'0} \cos{\Phi} +\epsilon R_{TT}^{y'0} \cos{2\Phi}) \nonumber
\end{eqnarray}
\vskip -0.6cm
\begin{eqnarray}
\label{ptran1}
\sigma_0 P_{\xi}'\!\!\! &=&\!\!\! {\cal{K}} ( c_- R_{LT'}^{\xi 0} \cos{\Phi} 
      + c_0 R_{TT'}^{\xi 0}),~~~~\xi=x'\!,z' \nonumber \\ 
\sigma_0 P_{y'}'\!\!\! &=&\!\!\! {\cal{K}} c_- R_{LT'}^{y'0} \sin{\Phi}
\end{eqnarray}
\noindent
Here $\sigma_0$ is the unpolarized cross section.  These definitions
can be related to the $(x,y,z)$ system via appropriate rotation operators.

The data were summed over all $\Phi$ angles to improve the statistical 
precision in the measurement.  In the summation, the induced components 
$P_{x',z'}^0$ and $P_{x,z}^0$, and the transferred components 
$P_{y'}'$ and $P_y'$, vanish identically.  In this 
case, the non-zero, helicity-gated yield asymmetries of eq.(\ref{asymmetry}) 
reduce to:
\begin{equation}
A_{\xi} = \alpha P_b \cos \theta_p^{RF} P'_{\xi}, \hskip 1.0cm 
\xi = \hat{i},\hat{k}, 
\end{equation}
\noindent
allowing for a direct extraction of $P'$ in a given kinematic bin with a 
linear fit of $A_{\xi}$ to $\cos \theta_p^{RF}$.  Note that different 
choices for the spin
axes lead to sensitivities of $P'$ to different subsets of the response 
functions.  The non-zero, $\Phi$-integrated transferred polarization 
components in the $(x',y',z')$ and $(x,y,z)$ systems are given by:
\begin{eqnarray}
P_{x'}' &=& c_1 R_{TT'}^{x'0} \hskip 0.75cm P_{z'}' = c_1 R_{TT'}^{z'0} \nonumber \\
P_x' &=& c_2 (R_{LT'}^{x'0} \cos{\theta_K^*} - R_{LT'}^{y'0} 
         + R_{LT'}^{z'0} \sin{\theta_K^*})  \nonumber \\
P_z' &=& c_1 (-R_{TT'}^{x'0} \sin{\theta_K^*} + R_{TT'}^{z' 0} \cos{\theta_K^*}).
\end{eqnarray}
\noindent
The normalization factors are given by $c_1 = c_0/K_0$ and 
$c_2 = c_-/(2K_0)$, where $K_0 = R_T^{00} + \epsilon_L R_L^{00}$.  This 
formalism is important to highlight as the hadrodynamic models provide the 
response functions in the $(x',y',z')$ system as their outputs.

\begin{figure}[htbp]
\vspace{4.2cm}
\includegraphics{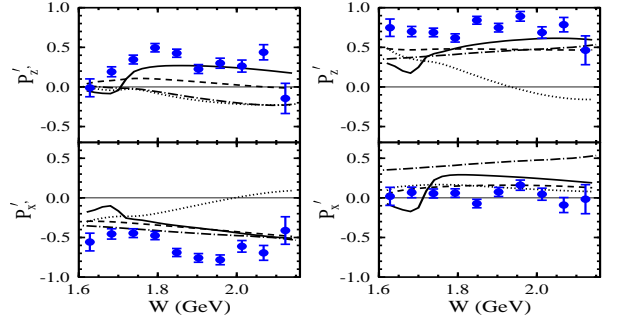}
\caption[]{Transferred $\Lambda$ polarization components $P_{z'}'$ and
$P_{x'}'$ (left) and $P_z'$ and $P_x'$ (right) at 2.567~GeV vs. $W$ (GeV) 
summed over all $Q^2$ and $d\Omega_K^*$.  Curves correspond to the hadrodynamic 
models: WJC92 (dotted), BM98 (dashed), BM02 (solid), J02 (dot-dash), averaged
over the experimental bins.}
\label{dpol1}
\end{figure}

\begin{figure}[htbp]
\vspace{7.7cm}
\includegraphics{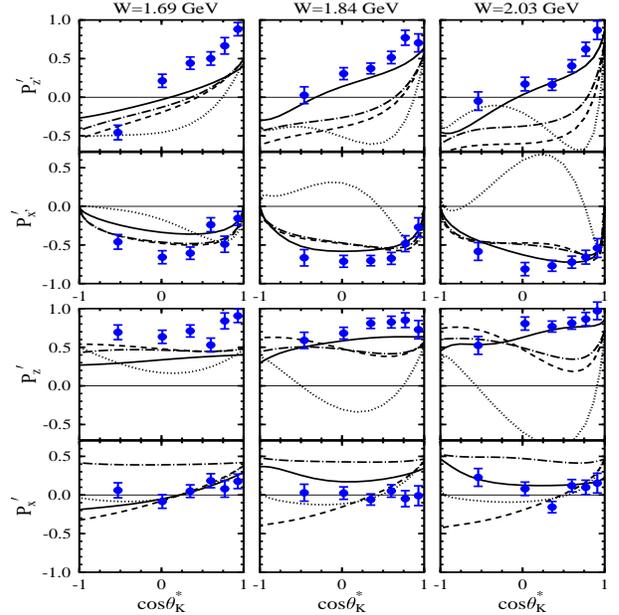}
\caption[]{Transferred $\Lambda$ polarization components $P_{z'}'$ and 
$P_{x'}'$ (upper) and $P_z'$ and $P_x'$ (lower) at 2.567~GeV vs. 
$\cos \theta_K^*$ summed over all $Q^2$ and $\Phi$ for three $W$ bins centered 
at 1.69, 1.84, and 2.03~GeV.  The curves are the same as Fig.~\ref{dpol1}.}
\label{dpol2}
\end{figure}

Our results are shown in Figs.~\ref{dpol1} and \ref{dpol2} compared to 
several hadrodynamic model calculations.  The error bars in these figures 
include statistical but not systematic uncertainties, which we estimate to 
be $\le$0.08 on the polarization~\cite{cnote}.  Fig.~\ref{dpol1} shows the 
$W$ dependence of $P'$ summed over all $Q^2$ and $d\Omega_K^*$ for our two 
choices of spin axes.  The data indicate sizeable $\Lambda$ polarizations 
with a relatively smooth variation with $W$.  The average polarization 
magnitude is roughly the same along the $z'$ and $x'$ axes, indicating equal 
strength in the $R_{TT'}^{z'0}$ and $R_{TT'}^{x'0}$ responses.  For the
other choice of axes, the polarization is maximal when projected along the 
$z$-axis (the virtual 
photon direction), while the component along the $x$-axis is consistent 
with zero.

Fig.~\ref{dpol2} shows the angular dependence of $P'$ summed over all $Q^2$ 
for three different $W$ bins from just above threshold to 2~GeV.  The 
polarization $P_{z'}'$ decreases with increasing $\theta_K^*$.  $P_{x'}'$ is 
constrained to be zero 
at $\cos \theta_K^*$ = $\pm$1 due to angular momentum conservation, and reaches
a minimum at $\theta_K^* \sim 90^{\circ}$.  Again, the maximum $\Lambda$ 
polarization occurs along the virtual photon direction.  This component, $P_z'$, 
is roughly constant with respect to $\cos \theta_K^*$ and $W$.  The component 
$P_x'$ in the electron-scattering plane is again consistent with zero.  The
components $P_{y'}'$ and $P_y'$ (not shown) are statistically consistent
with zero with respect to $W$ and $\cos \theta_K^*$ as expected~\cite{cnote}.

In each of the hadrodynamic models to which we compare our data, the coupling 
strengths have been determined by a simultaneous fit to the low-energy 
$K^- p \rightarrow \gamma Y$ radiative capture data and/or
$\gamma^{(*)} p \rightarrow K^+ Y$ data, by adding the non-resonant
Born terms with a number of resonances and leaving their coupling constants as 
free parameters bounded loosely by SU(3) predictions.  Different models have 
markedly different ingredients and fitted coupling constants.  In all cases, 
good agreement with the limited previous data has been achieved
\cite{wjc,mart,saghai,janssen}.  The comparison of the models to 
data can be used to provide indirect support for the existence of these excited 
states and their branching ratios into the strange channels.

\begin{table}[htpb]
\begin{center}
\begin{tabular} {|c|c|c|c|c|} \hline
Resonance              & WJC92 & BM98 & BM02 & J02 \\ \hline
$N^*(1650)$,$N^*(1710)$&  *    & *    & *    & *   \\ \hline 
$N^*(1720)$            &       &      & *    & *   \\ \hline
$N^*(1895)$            &       &      & *    & *   \\ \hline
$K^*(892)$             &  *    & *    & *    & *   \\ \hline
$K_1(1270)$            &  *    &      & *    & *   \\ \hline
$\Lambda^*(1405)$      &  *    &      &      &     \\ \hline
$\Lambda^*(1800)$,$\Lambda^*(1810)$&       &      &      & *   \\ \hline
\end{tabular}
\end{center}
\caption[]{Resonances included in the hadrodynamic models 
highlighted in this work included with the non-resonant Born terms.  References
to the models can be found in the text.}
\label{models}
\end{table}

Recent calculations have been guided by coupled-channels analyses
\cite{feuster,nstar} 
that recognize the importance of the $S_{11}$(1650), $P_{11}$(1710), and 
$P_{13}$(1720) $s$-channel resonances, which are also the only ones with a
known significant branching into the strange channels~\cite{pdg}.  The 
$p(e,e'K^+)\Lambda$ cross section data exhibit a forward peaking in 
$\theta_K^*$ that has been attributed to $t$-channel exchanges
\cite{hicks}.  For this reason the two lowest vector meson resonances 
$K^*$(892) and $K_1$(1270) are also typically included.  A comparison of the 
four different models employed in this work is included in Table~\ref{models}.  
These models were developed by Williams, Ji, and Cotanch (WJC92)~\cite{wjc}, 
Bennhold and Mart (BM02,BM98)~\cite{haber,mart}, and Janssen (J02)~\cite{janssen}. 
These models differ in their mix of $N^*$ resonances, e.g. BM02 and J02 both
include one of the missing quark-model states, the $D_{13}(1895)$.  Some of
the models include $Y^*$ resonances in the $u$ channel, and the most
recent models (all but WJC92) have included form factors at the hadronic 
vertices.  In this work, we have employed simple electromagnetic dipole form 
factors for the kaon and the hyperon.

The calculations generally do not reproduce the data in Figs.~\ref{dpol1} and 
\ref{dpol2}, however, the BM02 model best reflects the data.  It should be 
noted that the kink at $W \sim$1.7~GeV in the BM02 model implies a problem 
with modeling of the form factors of either the $P_{11}$(1710) or the 
$P_{13}$(1720) states.  While the comparison of the calculations to the data is 
illustrative to highlight the present deficiencies in the current models and 
their parameter values, the next step in the study of the reaction mechanism 
is to include our polarization data in the available database and 
to refit the set of coupling strengths. 

As noted earlier, our data reveal a simple phenomenology that indicates
the $\Lambda$ polarization is maximal along the virtual photon 
direction.  We note that the lack of a strong $W$ dependence is an 
indication that the data might be economically described in a flux-tube 
strong-decay framework.  In this picture we expect that the cross section is 
dominated by photo-absorption by a $u$ quark.  When viewed in the 
$\gamma^*$-$p$ Breit frame, after a $u$ quark has absorbed the virtual photon, 
there is an intermediate $u$-$(ud)$ system with the $u$ quark polarized along 
the photon direction (+$z$) due to the helicity-conserving vector interaction.  
Hadronization into the $K^+$-$\Lambda$ final state proceeds with the production 
of an $s\bar{s}$ pair that breaks the color flux-tube.  Because the $u$ quark 
hadronizes as a pseudoscalar $K^+$, the $\bar{s}$ quark spin is required to be 
opposite to that of the $u$ quark, i.e. in the -$z$ direction.  In the
non-relativistic quark model the entire spin of the $\Lambda$ is carried
by the $s$ quark (this assumption has been questioned in light of the
NMC ``spin crisis''~\cite{jaffe}).  Since we observe the $\Lambda$ polarization 
to be in the +$z$ direction, we conclude that the $s$ and $\bar{s}$ spins were 
anti-aligned when they were created, if the hadronization did not flip or rotate 
their spins. We note that the authors of Ref.~\cite{boros} also posit a two-step
process for the production of transversely polarized $\Lambda$ hyperons in the
exclusive $pp \to pK^+ \Lambda$ reaction, and come to the similar conclusion that
the $s$ and $\bar{s}$ quark pair must also have been produced with spins
anti-aligned.

A dominance of spin anti-alignment for the $s$ and $\bar{s}$ quarks is not 
consistent with the $S$=1 $^3\!P_0$ operator, which predicts a 2:1 mixture of
$s\bar{s}$ quarks produced with spins aligned vs. anti-aligned if the orbital
substates are equally populated.  Along with 
other observations of failure of the $^3\!P_0$ model (e.g. explaining 
$\pi_2 \to \rho \omega$ decay~\cite{barnes}),  the applicability of the 
$^3\!P_0$ model in describing all hadronic decays is brought into doubt.  
We await theoretical investigations on the effect of the functional form of the 
quark-pair-creation operator on hyperon polarizations when a single $s \bar{s}$ 
pair is produced.

We have reported the first double-polarization measurements in the resonance region 
for the $p(\vec{e},e' K^+)\vec{\Lambda}$ reaction.  Our data show a large degree of 
$\Lambda$ polarization, which is maximal along the virtual photon direction
(averaging $\sim$75\% for our kinematics). 
As this is the first polarization data set, inclusion into the available
database should make hadrodynamic models much more reliable 
for studies of missing-resonance production.  Additionally, we feel 
that a better handle on the form of the quark-pair creation operator will make 
baryon spectroscopic models more reliable, hence increasing our confidence in 
their predictions for missing resonances.

We would like to acknowledge the outstanding efforts of the JLab staff that made 
this experiment possible.  This work was supported in part by the the U.S. 
Department of Energy and National Science Foundation, the Istituto Nazionale di 
Fisica Nucleare, the French Centre National de la Recherche Scientifique,
the French Commissariat \`a l'Energie Atomique, and the Korea 
Science and Engineering Foundation.  The Southeastern Universities Research 
Association operates JLab for the U.S. Department of Energy under contract 
DE-AC05-84ER40150.

\end{document}